\documentclass[aps,prb,reprint,superscriptaddress,showpacs]{revtex4-1}

\usepackage{color}
\usepackage{graphicx}
\usepackage{amsmath}
\usepackage{amsfonts}
\usepackage{amssymb}
\usepackage{rotating}
\usepackage{hyperref}

\newcommand{\ket}[1]{|#1\rangle}
\newcommand{\braket}[2]{\langle #1|#2\rangle}

\allowdisplaybreaks

\usepackage{multirow}
\usepackage{braket}

\begin{document}

\author{Sebastian Wouters}
\email{sebastianwouters@gmail.com}
\affiliation{Center for Molecular Modelling, Ghent University, Technologiepark 903, 9052 Zwijnaarde, Belgium}
\author{Brecht Verstichel}
\affiliation{Department of Chemistry, Princeton University, Princeton, New Jersey 08544, USA}
\author{Dimitri Van Neck}
\affiliation{Center for Molecular Modelling, Ghent University, Technologiepark 903, 9052 Zwijnaarde, Belgium}
\author{Garnet Kin-Lic Chan}
\email{gkc1000@gmail.com}
\affiliation{Department of Chemistry, Princeton University, Princeton, New Jersey 08544, USA}

\title{Projector quantum Monte Carlo with matrix product states}

\begin{abstract}
We marry tensor network states (TNS) and projector quantum Monte Carlo (PMC) to overcome the high computational scaling of TNS and the sign problem of PMC. Using TNS as trial wavefunctions provides a route to systematically improve the sign structure and to eliminate the bias in fixed-node and constrained-path PMC. As a specific example, we describe phaseless auxiliary-field quantum Monte Carlo with matrix product states (MPS-AFQMC). MPS-AFQMC improves significantly on the DMRG ground-state energy. For the $J_1$-$J_2$ model on two-dimensional square lattices, we observe with MPS-AFQMC an order of magnitude reduction in the error for all couplings, compared to DMRG. The improvement is independent of walker bond dimension, and we therefore use bond dimension one for the walkers. The computational cost of MPS-AFQMC is then quadratic in the bond dimension of the trial wavefunction, which is lower than the cubic scaling of DMRG. The error due to the constrained-path bias is proportional to the variational error of the trial wavefunction. We show that for the $J_1$-$J_2$ model on two-dimensional square lattices, a linear extrapolation of the MPS-AFQMC energy with the discarded weight from the DMRG calculation allows to remove the constrained-path bias. Extensions to other tensor networks are briefly discussed.
\end{abstract}



\maketitle

\section{Introduction}

Tensor network states (TNS) and projector quantum Monte Carlo (PMC) are numerically exact methods for strongly correlated quantum states.\cite{schlwck_mps_rev, r_orus_rev,verstraete_rev, kalos_gfmc,kalos_book, bs_pmc, bss_afqmc, zhang_afqmc_cp, zhang_afqmc_pf, foulkes_rev_dmc,trivedi_lattice, fciqmc} TNS provide compact parametrizations of quantum states in terms of local tensors, and become exact with increasing bond dimension $D$.\cite{r_orus_rev,verstraete_rev,peps_initial, peps_pra,peps_prl,mera,branching_mera} Matrix product states (MPS), the basis of the density matrix renormalization group (DMRG),\cite{schlwck_mps_rev,white_prl,white_prb} are a widely successful example in one- and quasi-two-dimensional simulations. Although TNS provide an unbiased description of quantum states, they exhibit high computational scaling with $D$. For example, variational projected entangled pair states (PEPS) on a finite square lattice  exhibit $\mathcal{O}(\chi^2 D^8)$ scaling, with $\chi \geq D^2$ the virtual dimension trunctation in the approximate contraction,\cite{r_orus_rev,verstraete_rev, peps_initial,peps_pra,peps_prl} which limits practical applicability.

PMC encompasses multiple methods with the common feature that the ground state $\ket{\Psi_*}$ is obtained by stochastically applying a projector, such as $\hat{K}=e^{-\delta \tau \hat{H}}$, to an ensemble of walkers $\sum_k \ket{\phi_k}$.\cite{kalos_gfmc, bs_pmc, bss_afqmc, zhang_afqmc_cp,zhang_afqmc_pf, foulkes_rev_dmc, trivedi_lattice,PhysRevB.61.2599, j1j2_sorella, Boninsegni1996313,fciqmc} After sufficient applications, this ensemble stochastically represents the ground state $\ket{\Psi_*}$. For fermionic and frustrated systems, the walkers in PMC tend to represent both $\pm \ket{\Psi_*}$, leading to a vanishing signal-to-noise ratio for expectation values, the so-called {\it sign} problem.\cite{shiwei_chapter,foulkes_rev_dmc} One way to approach fermionic and frustrated systems is to use fixed-node (FN) or constrained-path (CP) approximations, which eliminate the sign problem by constraining walkers to a fixed phase relative to an approximate trial state $\ket{\Psi_T}$.\cite{anderson_fixed_node, zhang_afqmc_cp, van1994fixed} However, this introduces a bias. With an improved description of the nodal structure in FN approximations, or the nodal plane in CP approximations, the bias becomes smaller.\cite{10.1063_1.432868, ceperley_fn, 10.1063_1.463296, foulkes_rev_dmc, 10.1063_1.1604379, PhysRevB.69.125110, PhysRevLett.96.240402, PhysRevLett.Cyrus, doi10.1143JPSJ.76.084709, PhysRevB.89.125129} The bias can often not be easily removed as common choices of trial states, such as Jastrow-Slaters,\cite{foulkes_rev_dmc, 10.1063_1.1604379} cannot be easily improved.\cite{iqbal2013gapless} 

Here, we present a marriage of TNS and PMC that combines their strengths and removes their respective weaknesses. Specifically, we combine MPS with phaseless auxiliary field quantum Monte Carlo (AFQMC),\cite{zhang_afqmc_cp,zhang_afqmc_pf} yielding MPS-AFQMC, although the ideas extend equally well to other combinations of TNS and PMC.  MPS-AFQMC uses MPS as the trial state $\ket{\Psi_T}$ as well as to represent the walkers. This allows us to systematically remove the CP bias and improve the sign structure by increasing the trial bond dimension $D_T$, eliminating the main drawback of CP-PMC.

For the $J_1$-$J_2$ model on two-dimensional square lattices, we observe with MPS-AFQMC an order of magnitude reduction in the error for all couplings, compared to DMRG. The improvement is independent of walker bond dimension, and we therefore use bond dimension $D_W=1$ for the walkers. This leads to an $\mathcal{O}(D_T^2)$ computational scaling of MPS-AFQMC, which is lower than the $\mathcal{O}(D_T^3)$ scaling of the corresponding variational DMRG calculation. The high computational cost of TNS can therefore be addressed by a combined TNS-PMC approach. The error due to the CP bias is proportional to the variational error of the trial wavefunction. We show that for the $J_1$-$J_2$ model on two-dimensional square lattices, a linear extrapolation of the MPS-AFQMC energy with the discarded weight from the DMRG calculation allows to remove the constrained-path bias.

We note that in earlier work, TNS have been used with {\it variational} Monte Carlo.\cite{vidal_vmc_tns,cps,eps,vidal_vmc_mera, PhysRevB.85.045103, PhysRevLett.111.187205} However, this method can only stochastically reproduce the variational TNS energy of the ansatz under consideration, while PMC methods allow to improve on the variational ansatz.
We mention also valence-bond basis projector Monte Carlo methods, which similarly use walkers in a complicated valence-bond basis, but which so far have only been formulated for sign-free problems.\cite{evertz}

\section{Projector Monte Carlo}

We begin with a brief overview of PMC methods before proceeding to MPS-AFQMC. PMC, encompassing lattice and real-space diffusion Monte Carlo  (DMC), Green function Monte Carlo (GFMC), and auxiliary field quantum Monte Carlo (AFQMC), involves a choice of projector, walker basis, and FN or CP approximation.

Several PMC methods, including AFQMC, use the imaginary time propagator  
\begin{equation}
\hat{K}=e^{- \delta\tau \hat{H}}.
\end{equation}
The ground-state is obtained by
\begin{equation}
\ket{\Psi_*}=\lim\limits_{n \rightarrow \infty}(\hat{K})^n \ket{\Psi^{(0)}}, \quad \text{for }\braket{\Psi_*|\Psi^{(0)}}\neq 0.
\end{equation}
Each application of $\hat{K}$ is denoted a time step. $\ket{\Psi^{(n)}}$, obtained after $n$ time steps, is represented
by an ensemble of walkers
\begin{equation}
\ket{\Psi^{(n)}} = \sum_k \ket{\phi_k^{(n)}}.
\end{equation}
Observables are obtained as averages over the ensemble; for example the (mixed) estimator for the energy is\footnote{One of our referees pointed out the interesting connection of the mixed estimator in projector Monte Carlo to the concept of weak measurements in quantum computation.}
\begin{equation}
E^{(n)}_T  =  {\sum_k \braket{\Psi_T \mid \hat{H} \mid \phi^{(n)}_k}}/ {\sum_k \braket{\Psi_T \mid \phi_k^{(n)}}}.
\end{equation}
Common choices of walkers include real-space coordinates,\cite{DMC,foulkes_rev_dmc} product spin states,\cite{trivedi_lattice} and Slater determinants (SDs), \cite{zhang_afqmc_cp, zhang_afqmc_cp_prb, zhang_afqmc_pf} and we will later introduce MPS walkers. 
To apply $\hat{K}$ to the walkers, we first express it as a summation (integral) over a probability distribution function (PDF) $P(\mathbf{x})$ and operators $\hat{B}(\mathbf{x})$
\begin{align}
\hat{K} = \sum_{\mathbf{x}} P(\mathbf{x}) \hat{B}(\mathbf{x}), \label{eq:proppdf}
\end{align} 
where the choices of $P(\mathbf{x})$ and $\hat{B}(\mathbf{x})$ further differentiate the flavours of PMC. The only restriction in Eq.~(\ref{eq:proppdf}) is that $\hat{B}(\mathbf{x})$  maps a single walker to another walker of the same complexity: real-space coordinates should only change positions, SDs should remain SDs, or (in this work) the bond dimension of an MPS should not grow.

$\hat{K}$ is applied by sampling  $\mathbf{x}$ with probability $P(\mathbf{x})$, and updating the walker: 
\begin{equation}
\ket{\phi^{(n)}_k} = \hat{B}(\mathbf{x}) \ket{\phi^{(n-1)}_k}.
\end{equation}
A common way to improve statistics is to employ importance sampling with
respect to a trial state $\ket{\Psi_T}$. Then the
propagator is modified to 
\begin{equation}
\hat{K}_{\phi} = \sum_{\mathbf{x}} P(\mathbf{x}) 
\frac{\braket{\Psi_T | \hat{B}(\mathbf{x}) | \phi}}
{{\braket{\Psi_T | \phi }}} \hat{B}(\mathbf{x}) = N_{\phi} \sum_{\mathbf{x}} \widetilde{P}_{\phi}(\mathbf{x}) \hat{B}(\mathbf{x}),
\end{equation}
where $N_{\phi}$ is the normalization to turn $\widetilde{P}_{\phi}(\mathbf{x})$ into a PDF. The importance sampling propagator biases moves towards regions where the overlap with $\ket{\Psi_T}$ is large. The ensemble now consists of weighted walkers
\begin{equation}
\ket{\Psi^{(n)}} = \sum_k w_k^{(n)} \ket{\phi_k^{(n)}}.
\end{equation}
$N_\phi$ is accumulated into the weights
\begin{equation}
w_k^{(n)} = N_{\phi_k^{(n-1)}} w_k^{(n-1)},
\end{equation}
which are controlled via branching. If the walker weights are smaller than 0.25, or larger than 1.5, $\lfloor w_k^{(n)} + u \rfloor$ copies of the walker are kept with weight 1, with $u$ drawn from the uniform distribution on $[0,1[$. Note that this does not change the ensemble stochastically.
In the importance sampling representation, the state and mixed estimator for the energy are
\begin{eqnarray}
\ket{\Psi^{(n)}} & \propto & \sum_k w_k^{(n)} \ket{\phi_k^{(n)}} / \braket{\Psi_T \mid \phi_k^{(n)}}, \label{eq:newEnsemble} \\
E^{(n)}_T & = & \frac{\sum_k w_k^{(n)} \braket{\Psi_T \mid \hat{H} \mid \phi^{(n)}_k} / \braket{\Psi_T \mid \phi_k^{(n)}}}{\sum_k w_k^{(n)}}. \label{eq:projectedEnergy}
\end{eqnarray}

After sufficient time steps, the time-averaged ensemble stochastically represents $\ket{\Psi_*}$. Since only a limited number of walkers is used, the sampling
bypasses the exponential complexity for the representation of a quantum state.
The only issue is the sign (or phase) problem, i.e.
 $\pm \ket{\Psi_*}$ (or  generally, $e^{i\theta} \ket{\Psi_*}$ with $\theta \in \left[0,2\pi\right[$) are both fixed points of $\hat{K}$.
Define the nodal plane $\mathcal{N}_{*}$:\cite{zhang_afqmc_cp,zhang_afqmc_cp_prb,ceperley_fn}
\begin{equation}
\ket{\phi} \in \mathcal{N}_{*} \iff \braket{\Psi_{*} \mid \phi} = 0.
\end{equation}
If $\ket{\phi}$ can cross $\mathcal{N}_{*}$ to $-\ket{\phi}$ by successive application of the operators $\hat{B}(\mathbf{x})$
(which is the case  for general fermion and frustrated
spin propagators) then $\pm \ket{\phi}$ will occur with equal probability after infinite MC 
time. The signal representing $\ket{\Psi_*}$ then arises as a vanishing difference between  populations of walkers representing $\pm \ket{\Psi_*}$,
and estimators, such as the projected energy in Eq.~(\ref{eq:projectedEnergy}),
have large fluctuations from vanishing denominators $\braket{\Psi_T\mid\phi_k^{(n)}}$.

To recover a finite signal, we introduce the CP approximation. A trial wavefunction $\ket{\Psi_T}$ \cite{zhang_afqmc_cp,zhang_afqmc_cp_prb,foulkes_rev_dmc} constrains the walker paths to one side of the nodal plane $\mathcal{N}_{T}$, by rejecting moves which change the sign of the overlap with $\ket{\Psi_T}$. This completely eliminates the sign problem. However, if $\ket{\Psi_T}$ is not exact and $\mathcal{N}_{*}\neq \mathcal{N}_{T}$, 
this introduces a systematic bias, which is the main drawback of CP-PMC. 
This is now the only remaining error in mixed estimators such as Eq. \eqref{eq:projectedEnergy}.

\section{MPS-AFQMC}

We now turn to MPS-AFQMC, the subject of this work.
MPS (with open boundary conditions) are defined by
\begin{eqnarray}
\ket{\phi} = \sum\limits_{\{ s^z_i ; \alpha_j \}} A[1]^{s_1^z}_{\alpha_1} A[2]^{s_2^z}_{\alpha_1 ; \alpha_2} .. A[L]^{s_L^z}_{\alpha_{L-1}} \ket{s_1^z s_2^z .. s_L^z},
\label{eq:mps}
\end{eqnarray}
where the summation over each bond index $\alpha_j$ is truncated to $D$. By using MPS as the trial state (obtained in a prior variational DMRG calculation),
we can increase the trial bond dimension $D_T$ to improve the CP approximation.
This provides a systematic route to eliminate CP bias.
We can also use MPS as walkers. To see this, consider the AFQMC decomposition for $\hat{K}$ with a Hubbard-Stratonovich (HS) transformation.\cite{stratanovic} 
For concreteness, the spin Hamiltonian 
\begin{equation}
\hat{H} = \frac{1}{2} \sum_{ij} J_{ij} \hat{\vec{S}}_i . \hat{\vec{S}}_j + \sum_{i} h_i \hat{S}^z_i
\end{equation}
is studied, with $J_{ij} = \sum_k V_{ik} \gamma_k V_{jk}$ symmetric. 
The (non-unique) HS transformation rewrites $e^{-\delta \tau \hat{H}}$,  bilinear in the spin operators in the exponent,
in terms of propagators with an exponent linear in the spin operators. 
Defining  $\hat{v}^{w}_k = \sum_i \hat{S}^{w}_i V_{ik} \sqrt{- \gamma_k}$ with $w \in \{ x,y,z \}$ we have:
\begin{eqnarray}
\hat{H} & = & \sum_{i} h_i \hat{S}^z_i - \sum_{wk} \frac{\left(\hat{v}^{w}_k\right)^2}{2} = \hat{H}_1 - \sum_{wk} \frac{\left(\hat{v}^{w}_k\right)^2}{2},\\
e^{-\delta\tau \hat{H}} & = & \int d\mathbf{x} P(\mathbf{x}) \hat{B}(\mathbf{x}) + \mathcal{O}(\delta\tau^2), \label{propagatorHS}\\
\hat{B}(\mathbf{x}) & = & e^{- \delta\tau \hat{H}_1 / 2} e^{\sqrt{\delta \tau} \mathbf{x}.\hat{\mathbf{v}}} e^{- \delta\tau \hat{H}_1 / 2}, \label{eq:HSBx}\\
P(\mathbf{x}) & = & \frac{e^{-\mathbf{x}^2/2}}{(2\pi)^{3L/2}},
\end{eqnarray}
with $\hat{\mathbf{v}} = \left( \hat{v}_1^x,\hat{v}_1^y,\hat{v}_1^z,\hat{v}_2^x, ... \right)$ and $L$ the number of lattice sites. 
Since  $\hat{B}(\mathbf{x}) \equiv \prod_i \exp \left( \sum_w \alpha_i^w \hat{S}_i^w \right)$ is a product of {\it single-site} operators, applying
 $\hat{B}(\mathbf{x})$ to an MPS does not increase its bond dimension, allowing the use of MPS walkers.
The walker bond dimension $D_W$ can be smaller than $D_T$ (if $D_W=1$, the walkers are product states) and this significantly reduces computational cost, as discussed below. We have also studied other operators $\hat{K}$ and other decompositions \eqref{eq:proppdf},\cite{seba_thesis} but MPS-AFQMC was found to be the most promising variant.

The other aspects of MPS-AFQMC are formulated in precisely the same manner as standard 
phaseless AFQMC. For completeness, we briefly describe the phaseless CP approximation
introduced by Zhang.\cite{zhang_afqmc_pf} Because  $\hat{B}(\mathbf{x})$  in  Eq.~(\ref{eq:HSBx}) can be complex, the AFQMC sign problem appears as a phase problem.
The importance sampling propagator is implemented (up to $\mathcal{O}(\delta \tau^{3/2})$)~\cite{zhang_afqmc_pf}
as a biased diffusion process
\begin{align}
\hat{K}_{\phi} 
&=  \int d\mathbf{x} P(\mathbf{x}) \hat{B}(\mathbf{x} - \mathbf{y}_\phi) N_{\phi}(\mathbf{x},\mathbf{y}_\phi),
\end{align}
where $\mathbf{y}_{\phi}$ applies a constant force drift, and importance sampling is achieved
by choosing $\mathbf{y}_{\phi} =- \sqrt{\delta\tau} \frac{\braket{\Psi_T \mid \hat{\mathbf{v}} \mid \phi}}{\braket{\Psi_T \mid \phi}}$ 
which  minimizes  fluctuations in the normalization factor $N_{\phi}(\mathbf{x},\mathbf{y})$.
$N_{\phi}(\mathbf{x},\mathbf{y}_{\phi})$ 
 further takes the simple evocative form
\begin{eqnarray}
N_{\phi}(\mathbf{x},\mathbf{y}_{\phi}) & \approx \exp\left[ -\delta\tau \frac{\braket{\Psi_T \mid \hat{H} \mid \phi}}{\braket{\Psi_T \mid \phi}} \right] \approx \exp\left[ -\delta\tau E_L(\phi) \right], \label{weightFuncvtion}
\end{eqnarray}
where $E_L(\phi)=\Re \frac{\braket{\Psi_T \mid \hat{H} \mid \phi}}{\braket{\Psi_T \mid \phi}}$ is the local energy. The phaseless approximation is
imposed by forcing walkers to maintain a positive overlap
with the trial state, modifying their weights by
\begin{align}
w_k^{(n)}\to w_k^{(n)} \max(0, \cos(\Delta \theta)),
\end{align}
where $\Delta \theta$ is the phase of $\braket{\Psi_T|\phi_k^{(n)}}/\braket{\Psi_T|\phi_k^{(n-1)}}$. 
The quality of this nodal constraint depends on the quality of $\ket{\Psi_T}$, but as discussed above,  by using MPS as $\ket{\Psi_T}$, the error can be completely removed by increasing $D_T$.

The main cost of MPS-AFQMC comes from computing the local energies $E_L(\phi_k)$ at each time step. If  $\ket{\Psi_T}$ has
bond dimension $D_T$ and the walkers bond dimension $D_W$ ($D_W < D_T$), then this
only costs $\mathcal{O}(D_W D_T^2)$, lower than the $\mathcal{O}(D_T^3)$ associated with expectation values in a variational DMRG calculation. 
For the $J_1$-$J_2$-model on 2D square lattices, both MPS-AFQMC and DMRG scale as $\mathcal{O}(L^\frac{3}{2})$ in the system size, with $L$ the total number of lattice sites.

\section{Spin-$\frac{1}{2}$ $J_1$-$J_2$ model on 2D square lattices}

To demonstrate the power of this new MPS-AFQMC approach, we now apply it to calculate the
ground-state energies of the spin-$\frac{1}{2}$ $J_1$-$J_2$ model on two-dimensional square lattices of sizes $4\times 4$ and $6\times 6$ with periodic boundary conditions (PBC), and $8\times 8$ and $10\times 10$ with open boundary conditions (OBC). The $J_1$-$J_2$ model is defined by
\begin{equation}
\hat{H} = J_1\sum_{\langle ij\rangle}\mathbf{S}_i \cdot \mathbf{S}_j + J_2\sum_{\langle\langle ij\rangle\rangle}\mathbf{S}_i \cdot \mathbf{S}_j
\label{j1j2_ham}
\end{equation}
in which  $J_1$ is the coupling for nearest-neighbour spins, and $J_2$ the
coupling for next-to-nearest neighbour spins. 
The $J_1$-$J_2$ model is of fundamental interest because it is one of the simplest models with frustration.
For $J_2=0$, the model is the spin-$\frac{1}{2}$ Heisenberg model, whose
ground-state is gapless and unfrustrated, and when $J_2/J_1=1$, the ground-state displays collinear striped magnetic order. In between, calculations show an intermediate phase in the region 
$0.4 \lesssim J_2/J_1 \lesssim 0.62$ which appears gapped, and may be a $Z_2$ spin-liquid.\cite{j1j2_jiang,j1j2_sorella}



\begin{figure}
\centering
\includegraphics[scale=0.5]{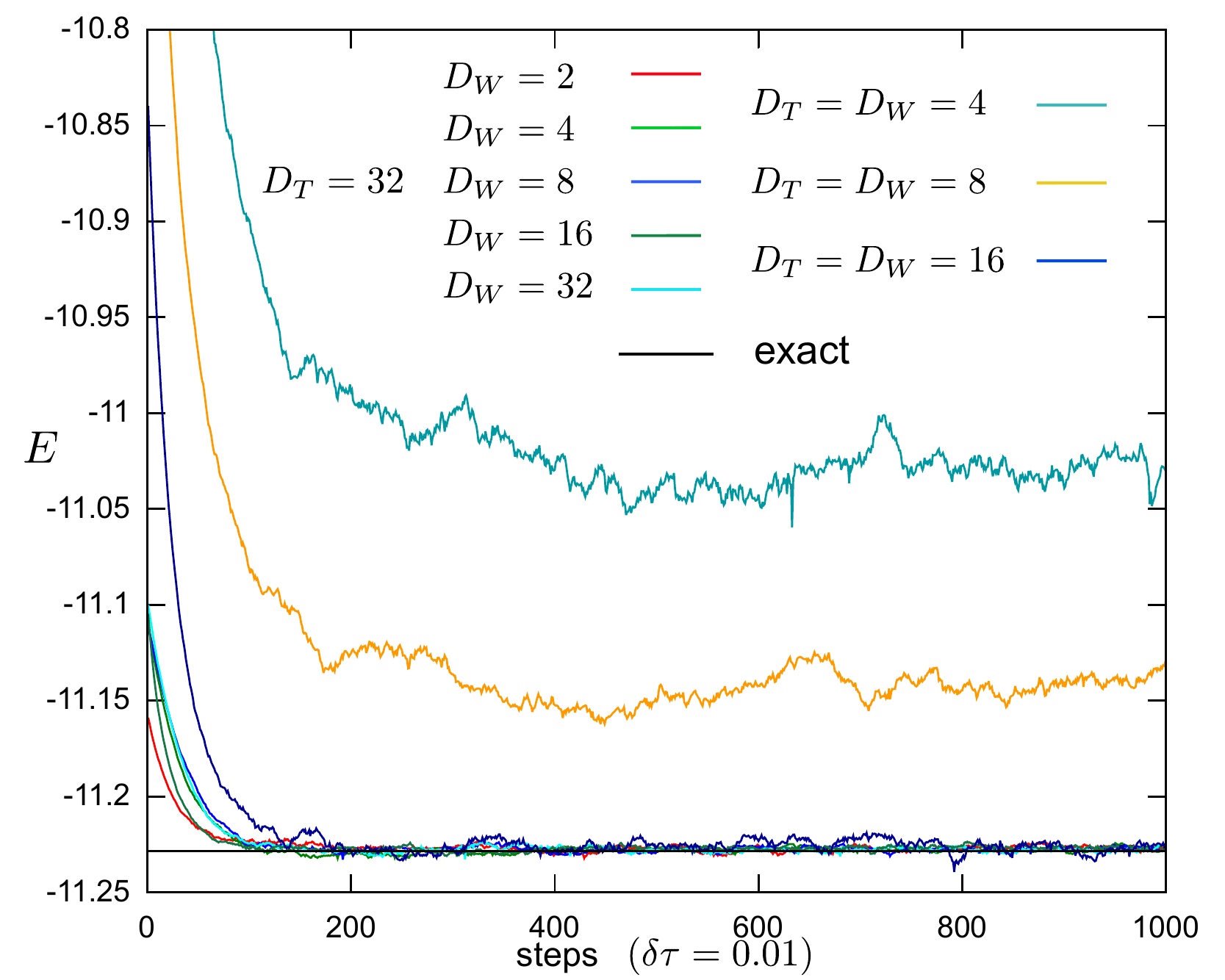}
\caption{\label{fig:walker_dim} Imaginary time evolution for a $4\times4$ $J_1$-$J_2$
 model with $J_2 = 0$ using different values for the trial bond dimension $D_T$ and
walker bond dimension $D_W$. }
\end{figure}

The calculations begin with variationally optimizing an MPS (using a DMRG code) with trial dimension $D_T$. Subsequently, we use MPS-AFQMC to propagate an ensemble of MPS walkers with bond dimensions $D_W$. All calculations have been performed with a time step of $\delta \tau = 0.01$, which was verified to yield a Trotter error within the range of the statistical error. We used sufficient time samples to obtain
statistical error bars below 0.01\% in the energy.
As a first check, we examine the dependence of the projected energy in Eq.~(\ref{eq:projectedEnergy})
on the walker dimension $D_W$.
Fig.~\ref{fig:walker_dim} shows the projected energy as a function
of imaginary time in the  smallest $4\times 4$ lattice with {$J_2/J_1=0.0$}, 
for a selection of $D_W$ and $D_T$. The zero-time energy ($y$ intercept) is close to the DMRG energy, and
the decrease of the curves shows the gain in accuracy using MPS-AFQMC. 
Interestingly, there appears to be {\it no} effect of walker bond dimension on either
the final MPS-AFQMC energy or its statistical fluctuations. Thus the quality of
the phaseless approximation depends only on $D_T$. We have therefore used walker bond dimension $D_W=1$ for the other calculations. 

\begin{figure}
\centering
\includegraphics[scale=0.48]{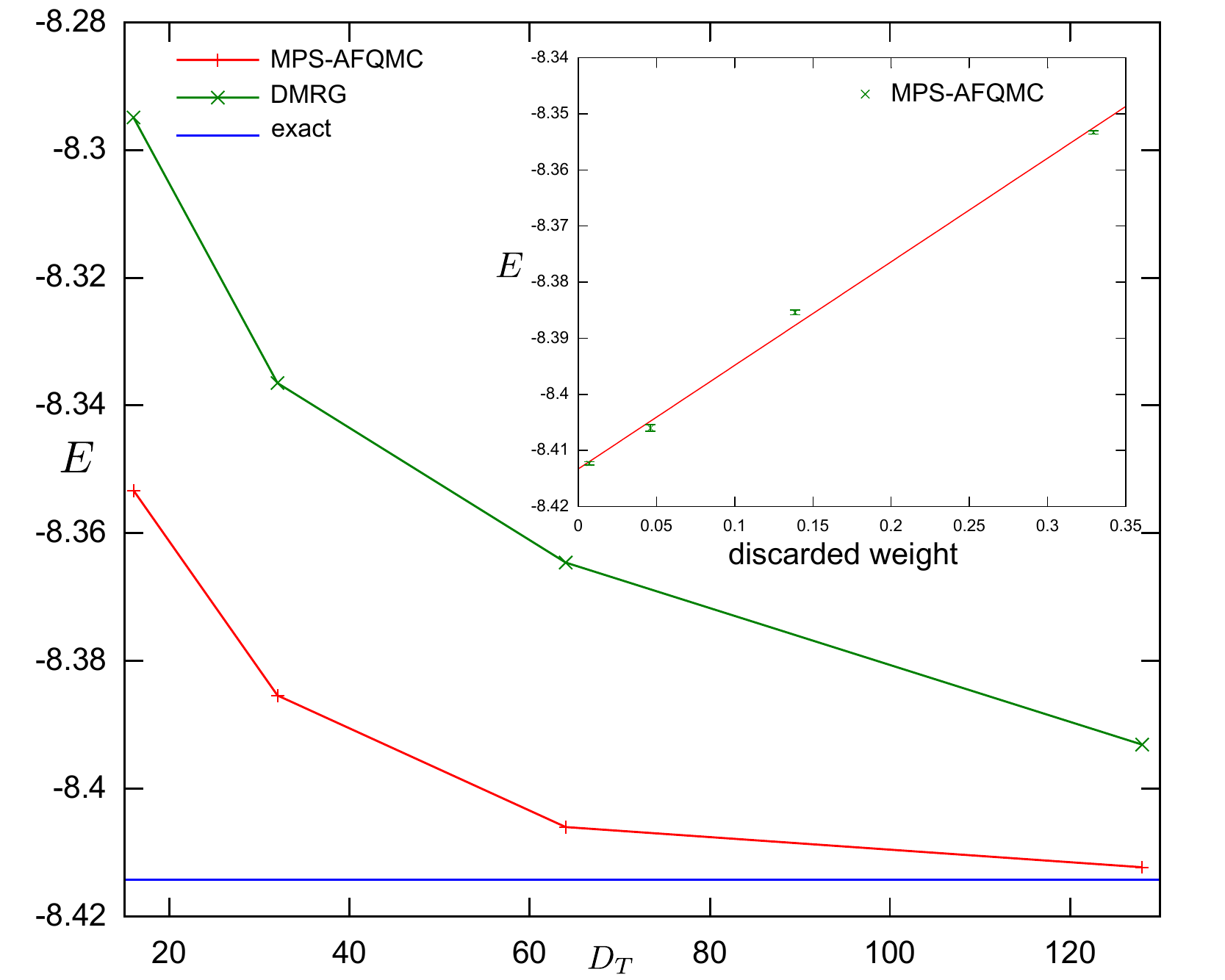}
\caption{\label{fig:energyM} Converged MPS-AFQMC and DMRG energies for different $D_T$
for a $4\times 4$ $J_1$-$J_2$ model (PBC) with {$J_2/J_1=0.6$}. 
 Inset: extrapolation of the MPS-AFQMC energy with respect
to discarded weight of the trial state.}
\end{figure}

As seen from Fig.~\ref{fig:walker_dim},  MPS-AFQMC provides a substantial
gain in accuracy over the initial DMRG energy. To see this more clearly, in Fig.~\ref{fig:energyM} we show
the converged MPS-AFQMC and DMRG energies for a variety of $D_T$ for the $4\times 4$ lattice, with {$J_2/J_1=0.6$}. 
 For this case, an MPS-AFQMC calculation with given $D_T$ reproduces the DMRG energy with a larger bond dimension of roughly  {$4D_T$}. Further, the convergence with $D_T$ is smooth for this system, resembling that of the DMRG energy itself. 
It is known that the DMRG energy can be extrapolated as a linear function of the discarded weight in the DMRG sweep algorithm.\cite{seba_thesis,chan_qcdmrg, chemps2}
Here, we obtain a similar linear dependence of the MPS-AFQMC energy with the DMRG discarded weight, as shown in the inset of Fig.~\ref{fig:energyM}. The CP error is therefore proportional to the variational error of the trial wavefunction.
This allows us to perform high quality extrapolations to $D_T=\infty$: in the case shown
 ($4 \times 4$ lattice, {$J_2/J_1=0.6$}), we obtain {$E(D_T=\infty)=-8.4133\pm0.0014$}, in accordance with the exact result {$-8.4143$}.


\begin{figure}
\centering
\includegraphics[scale=0.5]{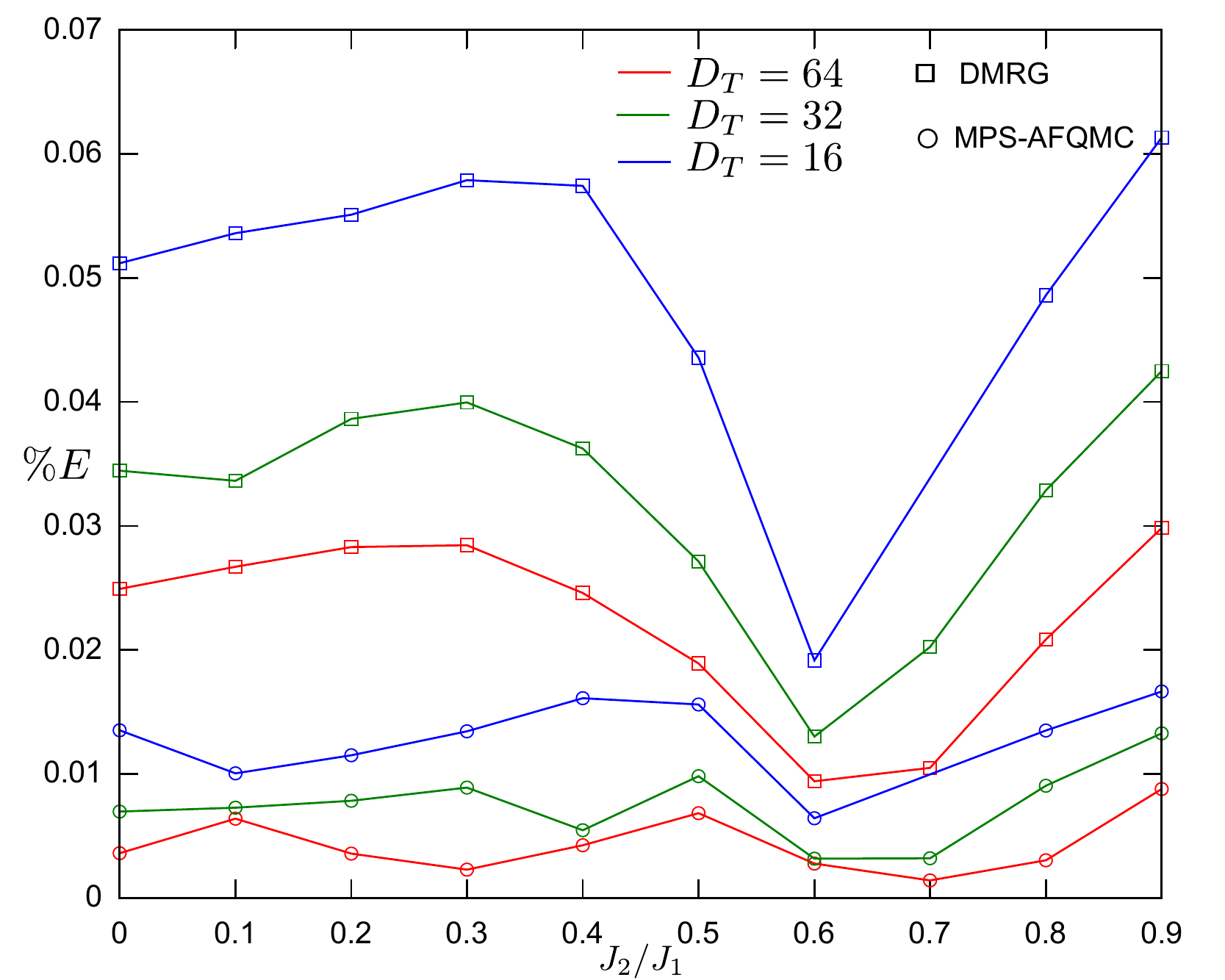}
\caption{\label{fig:6x6sum} Relative energy errors (from the exact result) for MPS-AFQMC and DMRG, as a function of $D_T$, for a $6\times6$ $J_1$-$J_2$ model (PBC) with $J_2$ ranging from 0 to $J_1$. } 
\end{figure}

\begin{figure}
\centering
\includegraphics[scale=0.5]{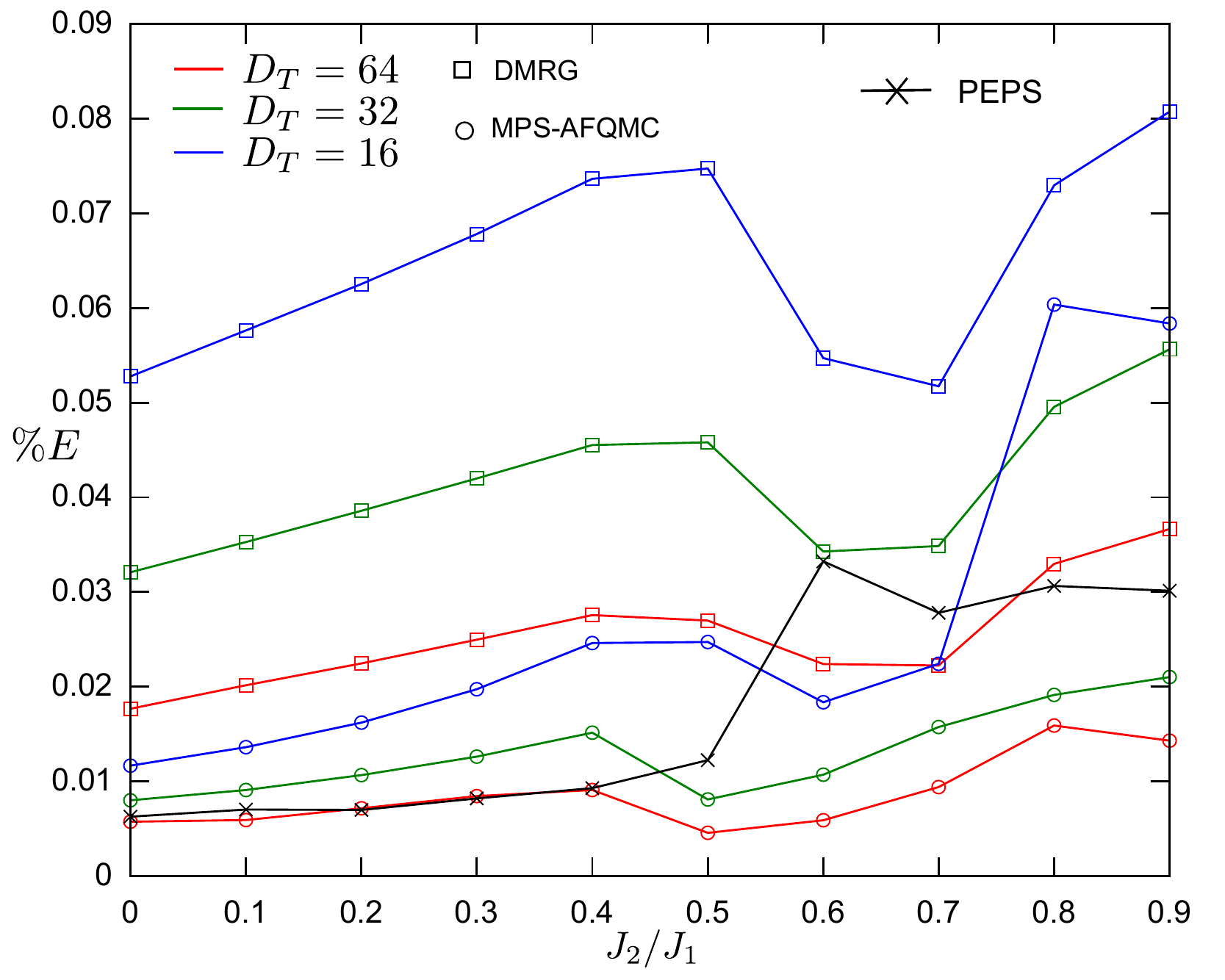}
\caption{\label{fig:8x8sum} Relative energy errors (from the exact result) for MPS-AFQMC and DMRG, as a function of $D_T$, for a $8\times8$ $J_1$-$J_2$ model (OBC) with $J_2$ ranging from 0 to $J_1$,  compared to PEPS results from Ref. \onlinecite{eps}.} 
\end{figure}

\begin{figure}
\centering
\includegraphics[scale=0.5]{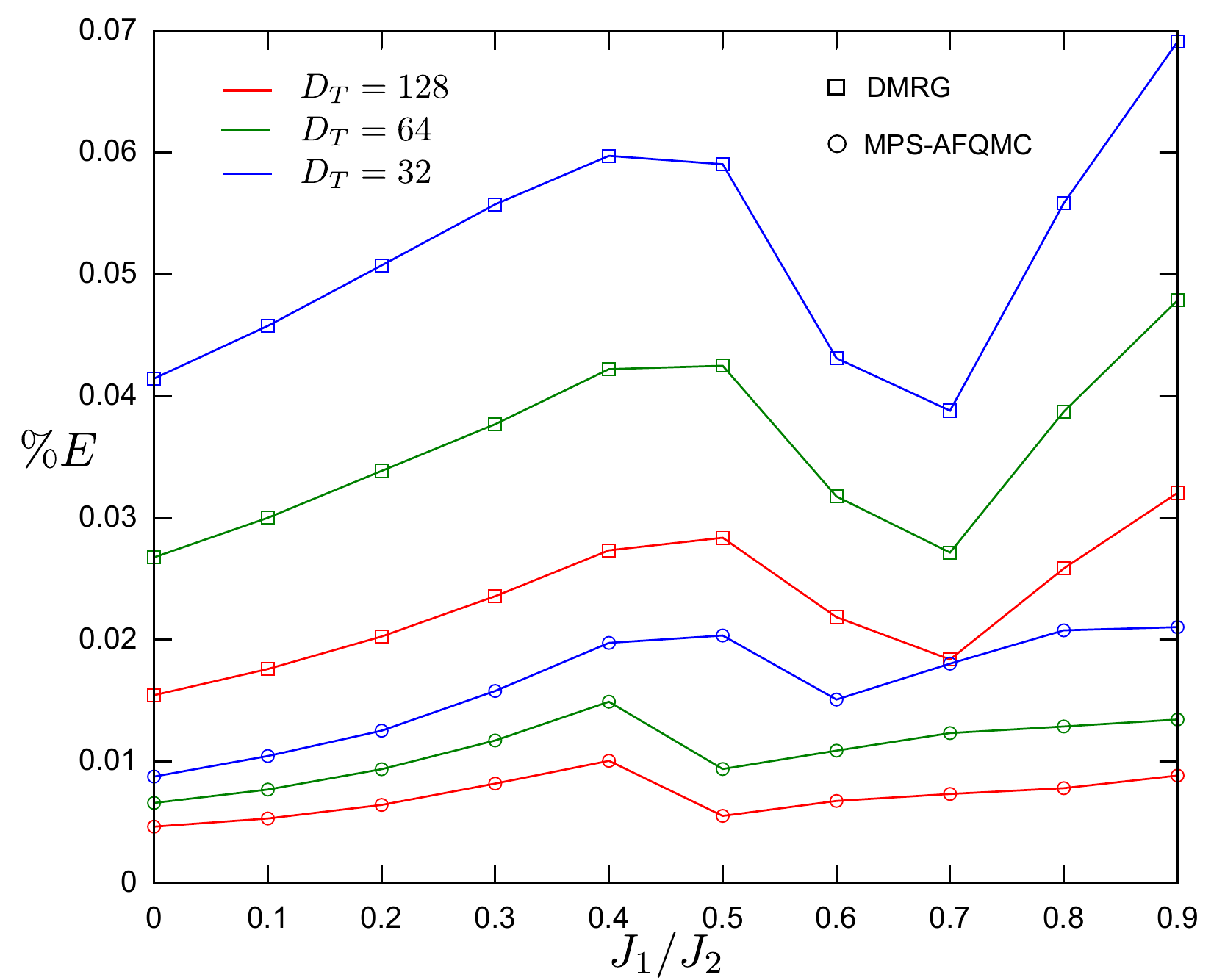}
\caption{\label{fig:10x10sum} Relative energy errors (from the exact result) for MPS-AFQMC and DMRG, as a function of $D_T$, for a $10\times10$ $J_1$-$J_2$ model (OBC) with $J_2$ ranging from 0 to $J_1$.} 
\end{figure}


We next examine the accuracy of MPS-AFQMC across different regimes of frustration by studying the energy as $J_2/J_1$ is varied. Figs.~\ref{fig:6x6sum}, \ref{fig:8x8sum} and \ref{fig:10x10sum} present the percentage error in the MPS-AFQMC and DMRG energies relative to exact results for the $6\times 6$, $8\times 8$ and $10\times 10$ lattices. The exact results were
obtained from spin-adapted DMRG calculations with $D_{\mathsf{SU(2)}}=2000$ reduced renormalized basis states. Across all frustration regimes, the MPS-AFQMC energy significantly improves on the DMRG energy, reducing the error by as much as {$80\%$}, even in the highly frustrated regime. Overall,
the MPS-AFQMC error tracks the modulations of the DMRG error as a function
of $J_2/J_1$, with the energies being more accurate in the gapped intermediate coupling regime than in the gapless $J_2/J_1=0$ and $J_2/J_1=1$ limits.

The MPS-AFQMC calculations, which did not use symmetries, required comparable time to a toy variational DMRG optimization of the trial state without symmetries for $D_T \approx 100$. In practice, we generated our trial states using our optimized spin-adapted DMRG code (with $\mathsf{SU(2)}$ symmetry). However, it is clear that for typical bond dimensions employed in DMRG ($D \approx 1000$) and typical  MPS-AFQMC parameters (100 walkers, 10000 time steps) PMC calculations will be highly competitive, if not faster, in timings, while achieving higher accuracy due to the effective bond dimension increase. 

For higher-order TNS, the reduction in computational complexity due to the single-layer structure ($D_W=1$) should be even more considerable. Further, PMC is highly parallel, in contrast to standard TNS optimization techniques.

When an MPS is used as a trial wavefunction for two-dimensional lattices, $D_T$ has to increase exponentially with lattice width to maintain a constant accuracy. For large lattices, it is therefore better to resort to PEPS to parameterize the trial wavefunction.

\section{Summary}

In conclusion, in this work we have described the marriage of tensor network states (TNS) and projector quantum Monte Carlo (PMC). The matrix product state auxiliary field quantum Monte Carlo (MPS-AFQMC) is a concrete realization of this marriage, which shows great promise. The use of an MPS trial wavefunction allows for a systematic removal of the CP error, which is the primary weakness of PMC methods in frustrated systems.

Further, the MPS-AFQMC method improves significantly on the variational DMRG ground-state energy, and does not depend on the bond dimension of the walkers. Product states ($D_W=1$) can therefore be chosen as walkers. This leads to a computational cost which scales only quadratically in the bond dimension of the trial wavefunction. The increase in MPS-AFQMC accuracy over DMRG can also be interpreted as an effective bond dimension increase. We demonstrated these improvements for the spin-$1/2$ $J_1$-$J_2$ model on the square lattice. In addition, we observed a linear dependence of the MPS-AFQMC energy with the DMRG discarded weight. The CP error is therefore proportional to the variational error of the trial wavefunction.

While we have only presented energies in this work, other observables and correlation functions can be obtained in MPS-AFQMC following standard PMC techniques.\cite{foulkes_rev_dmc} In addition, while we have discussed spin systems in this work, fermionic MPS allow for a direct extension to fermions, including long-range Hamiltonians such as the Coulomb interaction in {\it ab initio} DMRG.\cite{white_qcdmrg,chan_qcdmrg,legaza_qcdmrg, chan_rev, chemps2} Finally, an important next step will be to extend these ideas to higher dimensional tensor networks, such as projected entangled pair states,\cite{peps_initial,peps_pra,peps_prl} where the prohibitive computational scaling with bond dimension will be substantially reduced by PMC techniques, while providing  greater accuracy than the corresponding variational calculation for the same bond dimension.

During the revision process, we discovered Refs. \onlinecite{PhysRevB.62.14844} and \onlinecite{2014arXiv1404.2296C}. Ref. \onlinecite{PhysRevB.62.14844} provides an earlier combination of DMRG with a different kind of MC, lattice DMC, within the fixed-node approximation, and contains similar ideas to the current work. Ref. \onlinecite{2014arXiv1404.2296C}, which appeared after our submission, also considered the combination of MPS and tree TNS with lattice DMC, with results that are comparable to using lattice AFQMC.

\begin{acknowledgements}
S.W. received a Ph.D. fellowship and B.V. a postdoctoral fellowship from the Research Foundation Flanders (FWO Vlaanderen). 
The work was supported by the US National Science Foundation through grants OCI-1265278 and CHE-1265277.
\end{acknowledgements}

\bibliography{MPSQMC}

\end{document}